\documentclass[twocolumn,prl,twocolumn,superscriptaddress]{revtex4}
\usepackage{color}
\usepackage{amsmath}
\usepackage{amssymb}
\usepackage{graphicx}
\usepackage{tikz}
\usepackage{xr}
\usepackage{xc}

\usepackage{float}
\usepackage{bbm}

\usepackage{epsfig} 
\usepackage{verbatim}
\usepackage{array}
\usepackage{setspace}

\usepackage[unicode=true,
 bookmarks=true,bookmarksnumbered=false,bookmarksopen=false,
 breaklinks=false,pdfborder={0 0 1},backref=false,colorlinks=true]
 {hyperref}
\hypersetup{
 linkcolor=magenta, urlcolor=blue, citecolor=blue, pdfstartview={FitH}, hyperfootnotes=false, unicode=true}
 
\usepackage{color}

\makeatletter
\@ifundefined{textcolor}{}
{%
 \definecolor{BLACK}{gray}{0}
 \definecolor{WHITE}{gray}{1}
 \definecolor{RED}{rgb}{1,0,0}
 \definecolor{GREEN}{rgb}{0,1,0}
 \definecolor{BLUE}{rgb}{0,0,1}
 \definecolor{CYAN}{cmyk}{1,0,0,0}
 \definecolor{MAGENTA}{cmyk}{0,1,0,0}
 \definecolor{YELLOW}{cmyk}{0,0,1,0}
}

\usepackage{amsfonts}\usepackage{tabularx}\usepackage{dcolumn}\usepackage{bm}\usepackage{graphicx}\usepackage{epstopdf}

\hypersetup{urlcolor=blue}

\makeatother

\newcolumntype{C}[1]{>{\centering\arraybackslash$}p{#1}<{$}}

\usepackage{algorithm}
\usepackage{algorithmic}

\begin{document}

\widetext

\title{Low-Depth Optical Neural Networks}

\author{Xiao-Ming Zhang}
\affiliation{Department of Physics, City University of Hong Kong, Tat Chee Avenue, Kowloon, Hong Kong SAR, China}
\affiliation{Shenzhen Institute for Quantum Science and Engineering and Department of Physics,
Southern University of Science and Technology, Shenzhen 518055, China}
\author{Man-Hong Yung}
\email{yung@sustech.edu.cn}
\affiliation{Shenzhen Institute for Quantum Science and Engineering and Department of Physics,
Southern University of Science and Technology, Shenzhen 518055, China}
\affiliation{Shenzhen Key Laboratory of Quantum Science and Engineering, Southern University of Science and Technology, Shenzhen, 518055, China}
\affiliation{Central Research Institute, Huawei Technologies, Shenzhen, 518129, China}

\begin{abstract}
Optical neural network (ONN) is emerging as an attractive proposal for machine-learning applications, enabling high-speed computation with low-energy consumption. However, there are several challenges in applying ONN for industrial applications, including the realization of activation functions and maintaining stability. In particular, the stability of ONNs decrease with the circuit depth, limiting the scalability of the ONNs for practical uses. Here we demonstrate how to compress the circuit depth of ONN to scale only logarithmically, leading to an exponential gain in terms of noise robustness. Our low-depth (LD) ONN is based on an architecture, called Optical CompuTing Of dot-Product UnitS (OCTOPUS), which can also be applied individually as a linear perceptron for solving classification problems. Using the standard data set of Letter Recognition, we present numerical evidence showing that LD-ONN can exhibit a significant gain in noise robustness, compared with a previous ONN proposal based on singular-value decomposition [Nature Photonics 11, 441 (2017)]. 
 \end{abstract}
\maketitle

\section*{Introduction}
Photonic computation represents an emerging technology enabling high-speed information processing with low energy consumption~\cite{Flamini.18}. Such technology can potentially be applied to solve many problems of machine learning, which has already created a significant impact on the physics community~\cite{Biamonte.17,Mehta.18,Carleo.17,Carrasquilla.17,Ma.17,Bukov.18,Yang.18, Zhang.18,Gao.18}. In particular, efforts have been made for decades in developing optical neural networks (ONNs) with different approaches~\cite{Wagner.87, Jutamulia.96, Shen.17,Tait.17,Lin.18,Chang.18,Hughes.18,Bagherian.18,Penkovsky.19,Feldmann.19}. Recently, much progress has been made in developing scalable on-chip photonic circuits~\cite{Harris.17,Flamini.18,Wang.17,Carolan.15,Spring.13}, leading to a new avenue towards large-scale implementation of ONNs. Compared with its free-space counterpart, on-chip ONN has advantages in terms of programmability and integrability~\cite{Shen.17}. This unconventional hardware architecture could potentially revolutionize the field of AI computing.

In order to achieve scalable ONNs, various circuit designs have been proposed recently~\cite{Shen.17,Hughes.18,Bagherian.18}, and they share similar characteristics, such as the scaling complexity of the circuit depth and the form of the multiport interferometers. 
In particular, ONN-based deep learning has been experimentally demonstrated~\cite{Shen.17}, by applying singular-value decomposition (SVD) for constructing any given linear transformation. Physically, these unitary transformations can be achieved with multiport interferometers~\cite{Reck.94,Clements.16}, together with a set of diagonal attenuators.

However, the circuit structure of SVD-ONN is only applicable for linear transformation represented by a square matrix; with $N$-dimensional input and $M$-dimensional output of data, the SVD approach of ONN requires $O(\max(N,M))$ layers of interferometers. As each layer will introduce errors to its output, the scalability of the SVD approach of ONN is limited by the errors scaling as $O(\max(N,M))$. 
Moreover, for machine-learning tasks of practical interest, both cases $N\gg M$ (e.g. image recognition~\cite{He.16}) and $M\gg N$ (e.g. generative model~\cite{Goodfellow.14}) are very common. Therefore, the SVD approach would require appending lots of ancillary modes to ``square the matrix", increasing the spatial complexity of the ONN. 
 
To surmount the problem of robustness and flexibility, we propose an alternative approach of ONN for performing machine-learning tasks. Our ONN is constructed by connecting basic optical units, called Optical Computation of dot-Product Units (OCTOPUS), which optically outputs the dot-product of two vectors; the resulting circuit depth scales logarithmically~$O(\log N)$. Even a single OCTOPUS can be applied as an optical linear perceptron~\cite{Freund.99}. In addition, the noise robustness of the OCTOPUS exhibits an {\it exponential} advantage compared with the SVD approach (see~\cite{sm} for the theoretical analysis ). 

On the other hand, for constructing a deep neural network, we propose two variants of low depth ONN, called tree low depth (TLD) and recursive low depth (RLD) ONN. Both architectures involve OCTOPUS as basic optical computing units, and they are applicable for non-square transformation at each layer as well, i.e., $N \ne M$. The TLD-ONN requires fewer optical elements, but may cost more energy; the RLD-ONN involves a more complex structure, but it is more energy efficient. In terms of noise robustness, our numerical simulation suggests that TLD- and RTD-ONN have the same level of robustness, but both of them are significantly better than SVD-ONN.

\section*{Results}
\begin{figure} 
 \centering
\includegraphics[width=0.8\columnwidth]{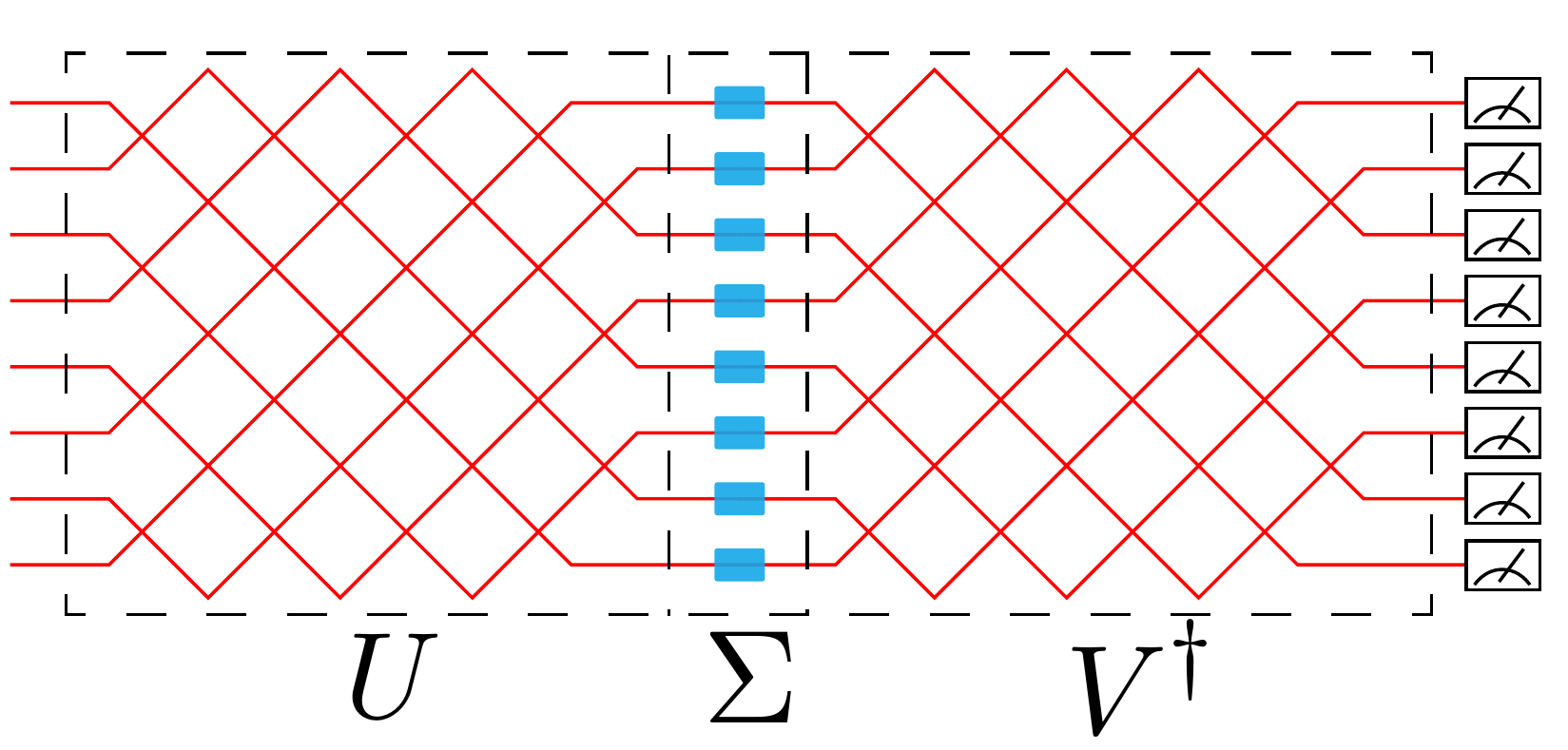}
\caption{SVD approach for linear transformation. The $8\times8$ square matrix is decomposed to two unitary $U$ and $V^{\dag}$, and a diagonal matrx $\Sigma$. The unitary matrices are realized by a set of M-Z interferometers, and $\Sigma$ is realized with a set of attenuators. }
\label{fig:svd}
\end{figure}

\textit{Linear Transformation---} Given a one dimensional real vector $\bm{x}$ and an $N\times M$ real transformation matrix $W$, our goal is to optically achieve the following linear transformation
\begin{equation}
\bm{y}=W\bm{x}.
\end{equation}
 In the SVD approach~\cite{Shen.17}, $N=M$ is assumed. Otherwise one needs to manually append  many ``$0$"s to square the corresponding matrix and vectors. Then, the matrix is decomposed as $W=U\Sigma V^{\dag}$ (see Fig.~\ref{fig:svd}), where $U$ and $V^\dag$ are unitary matrices, and $\Sigma$ is a diagonal matrix. In optical implementation, $U$ and $V^\dag$ can be realized with multiport interferometers of circuit depth $O(N)$, and $\Sigma$ can be realized with a set of attenuators or amplifiers~\cite{Shen.17,Steinbrecher.18}. 
 
\begin{figure}
\centering
\includegraphics[width=0.7\columnwidth]{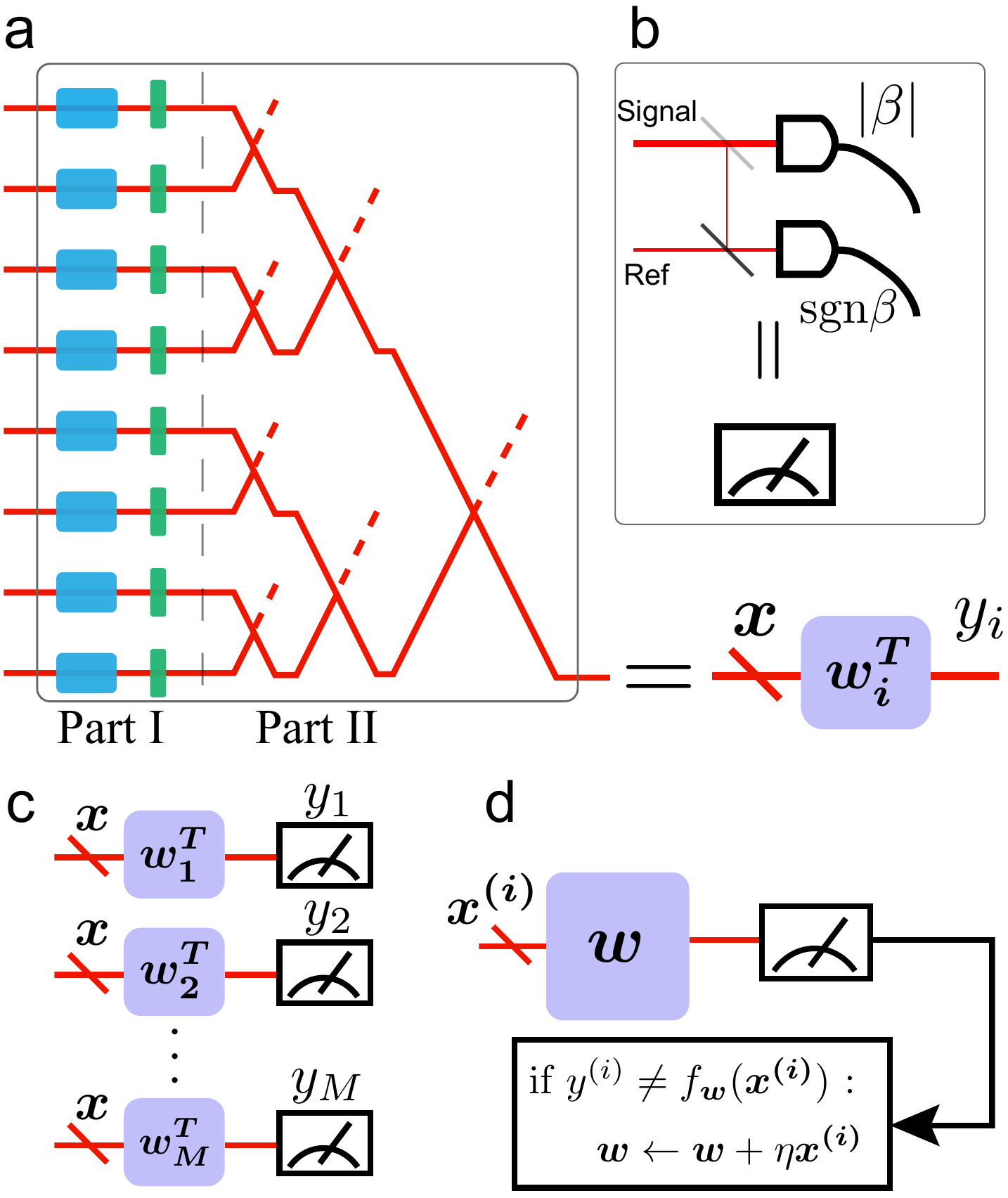}
\caption{\label{fig:orc} OCTOPUS, linear transformation and linear perceptron. \textbf{a} Sketch of the OCTOPUS calculating $y_i=\bm{w_i^T\cdot x}$. The dimension of input vector $\bm{x}$ is  $N=8$. Part I: attenuators (blue) and phase shifters (green) encode values of $\bm{w_i^T}$. Part II: interferometer tree performs the summation. \textbf{b} Amplitude measurement. Black and grey lines represent beam splitters. See Method of details. \textbf{c} Linear transformation with OCTOPUS. Each OPCTPUS corresponds to one row of the transformation matrix $\bm{w_{i}^T}$. \textbf{d} Sketch of the training process of optical linear perceptron. $\bm{x^{(i)}}$ and $y^{(i)}$ correspond to the training data and label at the $i$th iteration. $\bm{w}^T\bm{\cdot x^{(i)}}$ is calculated with OCTOPUS, after which we obtain the value of $f_{\bm{w}}(\bm{x^{(i)}})$. If $y^{(i)}\neq f_{\bm{w}}(\bm{x^{(i)}})$, the weight $\bm{w}$ is updated. }
\end{figure}
 
 In contrast, our OCTOPUS solves the same problem by calculating the elements of the output $\bm{y}$ ``one by one''. We require $M$ copies of the input $\bm{x}$, each of which serves as the input of one OCTOPUS. The $i$th OCTOPUS encodes the $i$th row of the matrix $W$ (denoted with $w_i$), and aims at calculating the $i$th element of the result $y_i=\bm{w_i}^T \cdot \bm{x}$.

The structure of OCTOPUS is shown in Fig.~\ref{fig:orc}a (see Methods for details).  
 At Part I, $\bm{w_i}^T$ is encoded with a set of tunable attenuators and phase shifters. In particular, the attenuators encode the magnitude of $\bm{w_i}^T$, while the phase shifters conditionally add a $\pi$ phase to the signal when the elements are negative. At Part II, we require a set of optical Hadamard transformation~\cite{Reck.94,Clements.16}. After each Hadamard transformation, we only trace the output port corresponding to the ``sum" of its input (other paths denoted with dash lines are discarded). They are constructed as an interferometer tree of $n$ layers with totally $N=2^n$ input ports and $1$ output port. The amplitude of the final output becomes $\frac{1}{\sqrt{N}}\bm{w_i}^T \cdot \bm{x}=\frac{1}{\sqrt{N}}y_i$, which is the desired outcome multiplying a constant. 

Note that the circuit depths of the SVD approach and the OCTOPUS approach are very different. The former scales linearly $O(N)$ and the latter logarithmically $O(\log N)$, leading to a dramatic difference in terms of the noise robustness against encoding errors of the optical elements.
Specifically, let us denote the output vector subject to noise by $\bm{\widetilde{y}}$. The error can be quantified by \textit{cosine distance}
\begin{equation}
\mathcal{D}(\bm{y},\bm{\tilde{y}})\equiv1 - \frac{\bm{y}\cdot\bm{\tilde{y}}}{\|\bm{y} \|\|\bm{\tilde{y}} \|},
\end{equation}
which has been widely adopted in classification problems~\cite{Nair.10,Nguyen.10,Dehak.11}.

\begin{figure}
\centering
\includegraphics[width=0.9\columnwidth]{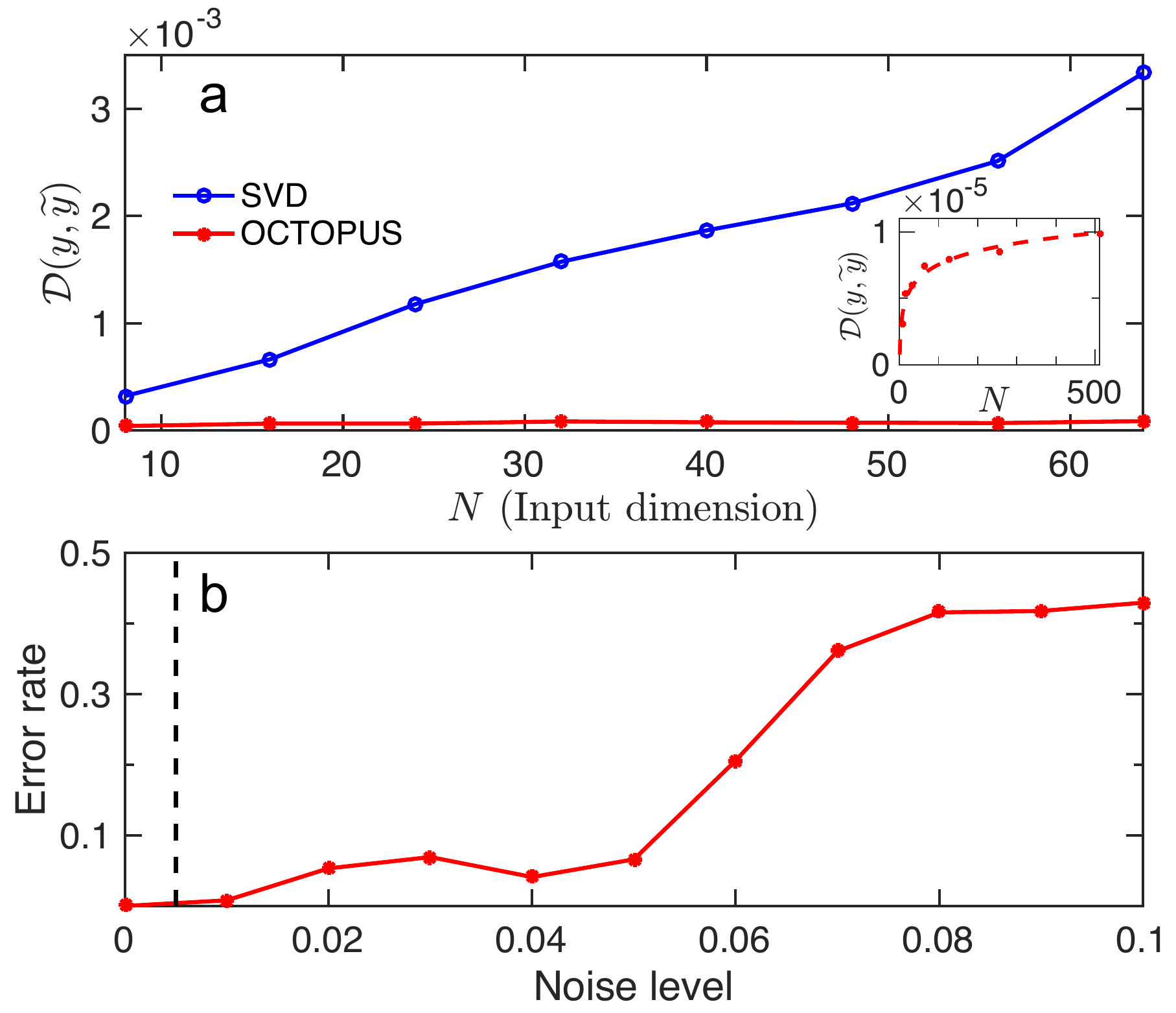}
\caption{\label{fig:com} \textbf{a} Robustness of the optical linear transformation. We let the dimension of input and output vectors to be identical, i.e., $M=N$. Main panel: comparison of the cosine distance $\mathcal{D}(\bm{y},\bm{\widetilde{y}})$ for SVD and OCTOPUS approaches. Encoding error level is set to be $\sigma_I=\sigma_A=0.005$. Inset: $\mathcal{D}(\bm{y},\bm{\widetilde{y}})$ for OCTOPUS approach when $\sigma_I=0.005$, $\sigma_A=0$. Dots are simulation data, dash lines are fitting with $\mathcal{D}(\bm{y},\bm{\widetilde{y}})=A\log N+B$. All results are averaged over 10 runs. \textbf{b} Linear perceptron simulation on the ``Iris'' data set. Red dots correspond to the error rate versus noise level $\sigma=\sigma_I=\sigma_A$ after $1000$ iterations of training. The red line is the guide for the eye. Black dash line corresponds to $\sigma=0.005$. Results are averaged over 100 runs. }
\end{figure}

\begin{figure} 
\centering
\includegraphics[width=0.9\columnwidth]{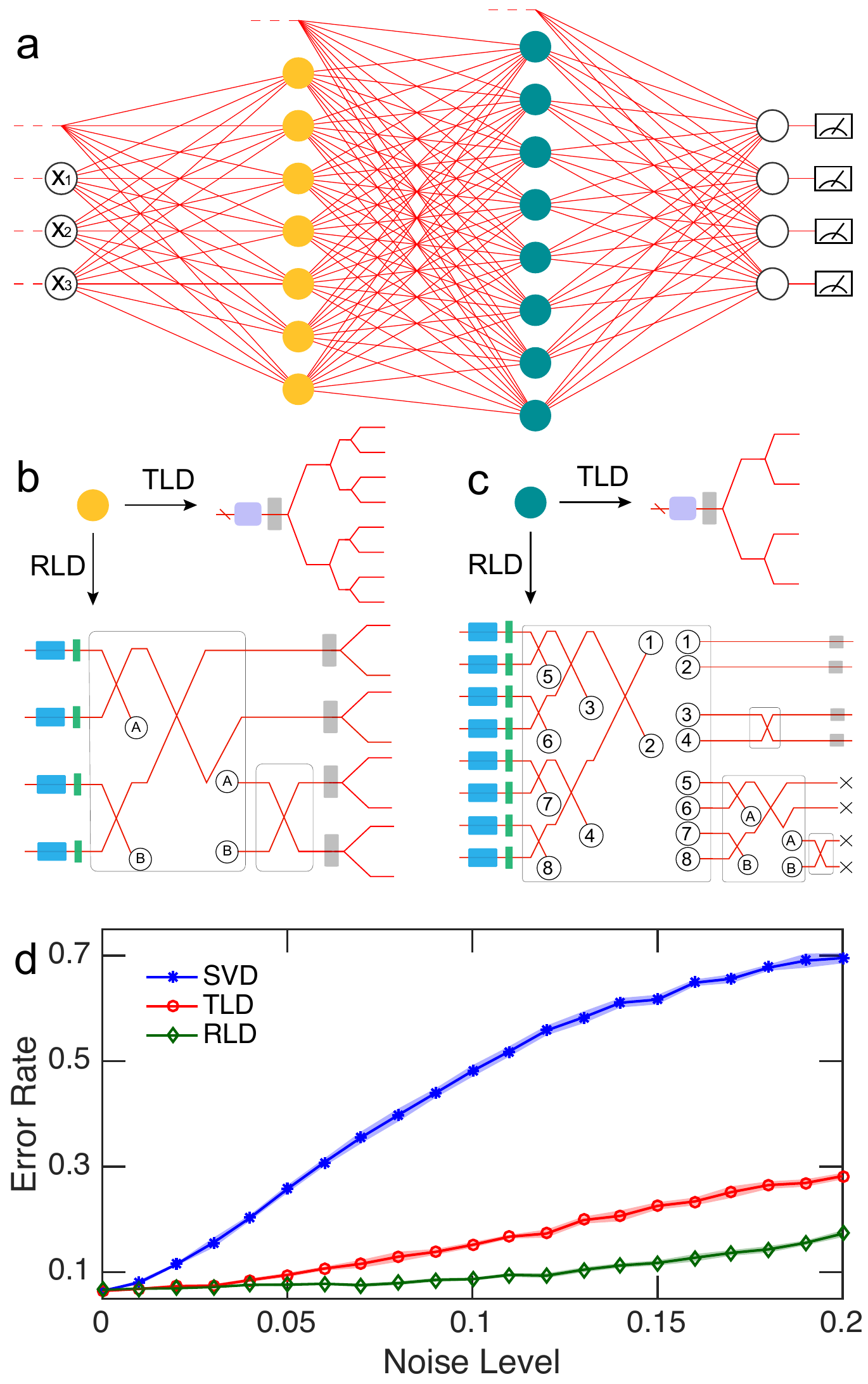}
\caption{\label{fig:net} \textbf{a} General structure of TLD- and RLD-ONN with $3\times7\times8\times4$ neurons. Red lines represent the optical paths. The input data denoted with $\bm{x}=[x_1,x_2,x_3]$ is encoded at the first layer. Neurons are represented with circles. \textbf{b} Realization of the neurons at the second layer of the network at \textbf{a}. For TLD-ONN, the input signal first passes through an OCTOPUS and then a nonlinear activation. Finally, it is split uniformly into several paths. For RLD-ONN, the input signals pass through sets of tunable attenuators (blue) and phase shifters (green). Then, several interferometer trees are appended recursively, until all output ports are connected to all input ports. Here, paths with same label (``A" or ``B") are connected to each other.   The signals then pass through sets of nonlinear activation, after which each path is splitted into two. \textbf{c} Realization of the neurons at the third layer. For RLD-ONN, at the end of the transformation, four paths denoted with black cross are discarded. \textbf{d} Error rate comparison of the Letter Classification task. The color regime corresponds to the confidential interval of $99\%$ for the error rates. More details are provided in Methods section.}
\end{figure}

Our simulation results comparing the SVD approach with OCTOPUS against Gaussian noises are shown in Fig.~\ref{fig:com}a. For the SVD approach, the error  $\mathcal{D}(\bm{y},\bm{\widetilde{y}})$ increases linearly with $N$. On the other hand, the error for OCTOPUS grows only very slowly. These results are consistent with the scaling of the circuit depths of the two approaches. In Supplemental Material~\cite{sm}, we provide further theoretical analysis on the noise effect. For SVD approach, the error scales linearly with the data size $N$, $\mathcal{D}(\bm{y},\bm{\widetilde{y}})\sim \sigma_I^2 N +\sigma_A^2$, where $\sigma_I$ and $\sigma_A$ represent noise level for interferometers and attenuators respectively. However, for OCTOPUS, the error scales only logarithmically, $\mathcal{D}(\bm{y},\bm{\widetilde{y}})\sim  \sigma_I^2 \log{N} +\sigma_A$. This exponential advantages of the OCTOPUS approach agrees well with our numerical results (see Fig.~\ref{fig:com}a). 

\textit{Linear perceptron---}  As OCTOPUS enables one to address each element of the input vector directly, it is possible to optically realize linear perceptrons~\cite{Rosenblatt.57,Mackay.03} with OCTOPUS for solving classification problems in machine learning.

For the binary case, the goal of linear perceptron is to output a hyperplane separating two classes of data labeled by either $0$ or $1$, allowing us to make prediction on the unlabelled new data.  
More precisely, with a set of training data $\{ \bm{x} \}$, one needs to determine the parameters $\bm{w}$ for the following function $f_{\bm{w}}(\bm{x})$:
\begin{eqnarray}
f_{\bm{w}}(\bm{x})= \left\{
\begin{array}{rcl}
1      &    & \bm{w}\bm{\cdot x}\geqslant0 \\
0    &  & \bm{w}\bm{\cdot x}<0   \label{eq:percep}\\  
\end{array} \right. \ ,
\end{eqnarray}
which can be realized with a single use of OCTOPUS followed by an appropriate measurement. As shown in Fig.~\ref{fig:orc}d, at the $i$th iteration, we use training data $\bm{x^{(i)}}$ as the input, determining whether its corresponding label $y^{(i)}=f_{\bm{w}}(\bm{x^{(i)}})$. If it is not, the weight is updated according to $\bm{w}\leftarrow \bm{w}+\eta \bm{x^{(i)}}$, where $\eta$ is the learning rate.  

Fig.~\ref{fig:com}b shows the simulation results of our optical linear perceptron with ``Iris Data Set''~\cite{Blake.98}. We define error rate as the rate of providing incorrect prediction on the label. Remarkably, the error rate remains under $0.1$ when the noise level $\sigma\leqslant0.05$. 

\textit{Low-depth ONN (LD-ONN)---} It is well known that linear perceptron performs well with relatively simple tasks. However, for problems involving complicated non-linear relations, one may consider deep neural networks. In the following, we present two variants of multi-layered ONN, namely, Tree Low-Depth (TLD) and Recursive Low-Depth (RLD) ONN. Both of them share a similar structure, as illustrated in Fig.~\ref{fig:net}a. Again, the input  data $\bm{x}$ is encoded at the first layer containing an attenuator and a phase shifter at each node. Optical computation is performed at  each neuron (denoted with colored circle), encapsulating the trainable parameters of the networks. Furthermore, the optical paths at the top of each layer represent the ``biases" of the corresponding layer.

As shown in Fig.~\ref{fig:net}b,c, for TLD-ONN, each neuron consists of an OCTOPUS together with a nonlinear activation function, which can be physically realized with non-linear crystal~\cite{Shen.17}, measurement~\cite{Hughes.18} or optical amplifier~\cite{Connelly.07,sm}. Then, each path is distributed uniformly to many paths, which are the inputs of the neurons at the next layer. More details are given in the Methods section and Supplemental Material~\cite{sm}.

Note that in TLD-ONN, the OCTOPUS only picks one path as its output; many paths are discarded. To realize a deep ONN, one may need a strong light source or amplify the signal at each layer. Alternatively, we may re-structure the ONN, which is the motivation for developing the RLD-ONN approach. 

As shown in Fig.~\ref{fig:net}b and c, for RLD-ONN, the input signals first pass through a set of trainable attenuators and phase shifters, followed by a ($3$ or $2$-layer) interferometer tree. Different from TLD-ONN, no signal are discarded after these steps. At this point, only two paths (such as ``$1$" and ``$2$" in Fig.~\ref{fig:net}c) are fully connected to the corresponding input ports. In order to connect all other output paths, the remaining paths are sent to interferometer trees with smaller size recursively. Then, the nonlinear activation is applied to all output paths. If the number of neurons at the next layer is larger than the current number of output paths, the output paths can be expanded with $50/50$ beam splitter (Fig.~\ref{fig:net}b); if it is less, one can just discard several output paths (Fig.~\ref{fig:net}c)\cite{note_1}. Note that the way of connecting input and output paths are not unique, so further optimization can be performed. Although RLD-ONN requires more optical elements, the circuit depth remains to be logarithmic.

The universality approximation theorem states that a feed-forward neural network with as few as a one hidden layer can approximate any continuous function to an arbitrary accuracy (under some mild assumptions on the activation function)~\cite{Hornik.91}, which is the foundation of neural computation. In Methods section, we show that the transformation of TLD-ONN is equivalent to standard feed-forward nerual network. And in~\cite{sm} we show that for any given from of one-hidden-layer TLD-ONN, there always exist a RLD-ONN that is equivalent to it. An illustration is also provided in Fig.~S2. Therefore, both TLD- and RLD-ONN proposed in this work are universal.

To compare the performance of LD-ONNs with SVD-ONN~\cite{Shen.17}, we perform numerical simulation on the  ``Letter Recognition'' data set~\cite{Blake.98}, classifying letters ``A", ``B", ``C" and ``D" (see Methods section for technical details). The ONN used in the simulation contains one hidden layer with $64$ neurons. While the SVD- and TLD-ONN are trained with standard back-propagation method, the RLD-ONN is trained with ``forward propagation"\cite{Shen.17}. In this work, we consider the training as a pre-processing, i.e., the parameters of the network are first trained in conventional computer. However, the training can also be realized optically with little assistance from electronic devices~\cite{sm}. As shown in Fig.~\ref{fig:net}d, when noise level $\sigma=0$, the error rates for SVD-, TLD- and RLD-ONN are $6.4\%$, $6.4\%$, $6.6\%$ respectively. These values may be improved by further optimizing the hyperparameters. As the noise level increases, TLD- and RLD-ONN have comparable error rates, but both of them are significantly lower than the error rate of SVD-ONN. 

\section*{Discussion}
 A summary of SVD-, TLD- and RLD-ONN is given in Table.~\ref{tab:com}, providing a comparison of the cost for an ONN layer with input dimension $N$ and output dimension $M$. Both the TLD- and RLD-ONN have logarithmic circuit depth, leading to exponential improvements on the error scalings compared with the SVD approach. Furthermore, TLD-ONN requires less number of optical elements, but we note that it also requires discarding more paths during the implementation.  On the other hand, RLD-ONN requires discarding much fewer paths (same as SVD-ONN), but at the cost of a larger number of optical elements. 

\begin{table}
\centering
\caption{\label{tab:com} Summary of the cost per layer for different ONN structures. All values corresponds to the order, $O(\cdot)$, for realizing one layer of the neural network with input dimension $N$ and output dimension $M$.}
\medskip
\begin{tabular}{lccc}
\hline
&SVD & TLD &RLD\\ 
\hline
Circuit depth& $\max(N,M)$&$\log N$&$\log N$\\
Error scaling &$ \max(N,M)$&$\log N$&$\log N$ \\
Number of elements &$\max(N^2,M^2)$& $NM$&$N^2M$\\
\hline
\end{tabular}
\end{table}

Note that several simplifications can further be made on the LD-ONN structures. Firstly, as discussed in Ref~\cite{sm}, the ONN can be binarized with amplifier working at the saturation regime. With the binarization of the weight, biases and activation function, the attenuators (for magnitude encoding) at each OCTOPUS can be removed. Secondly, instead of encoding the parameters at the phase shifter and attenuators, they can also be encoded at the interferometers. More specifically, it is possible to remove the part I of OCTOPUS, and replace the Hadamard transformation at Part II by tunable interferometers. Thirdly, after the training process, the network can be ``compressed" with the ``pruning" technique\cite{Han.15}: removing all paths with weights below a threshold. Above improvements or revision could helps reducing the complexity of the hard-ware architectures. 

To conclude, based on OCTOPUS, we present a new architecture of ONN for machine learning, which provides exponential improvements on the robustness against encoding error. We discussed different schemes of optical linear transformation, linear perceptron, and two variants of multi-layered ONNs. Numerical simulations with random transformations and standard machine learning data sets are employed to justify the robustness of our schemes. The proposed LD-ONN can be directly implemented with current photonic circuits\cite{Flamini.18} technology. Our proposal, combined with appropriate realization of nonlinear activation (For example, the scheme in Ref~\cite{Williamson.19}), provides a possible solution to solving machine-learning tasks of industrial interest with robust, scalable, and flexible ONNs.

\section*{Methods}

\textbf{OCTOPUS.}
We consider two vectors $\bm{\alpha}=\left[\alpha_1,\alpha_2,\cdots,\alpha_N \right]^T$ and $\bm{\gamma}=\left[\gamma,\gamma_2,\cdots,\gamma_N\right]^T$ with $\alpha_i\in \mathbb{R}$ and $\gamma_i\in[-1,1]$, and assume $N=2^n$ with $n\in \mathbb{Z}^+$. They corresponds to $\bm{x}$ and $\bm{w_i^T}$ in Fig.~\ref{fig:orc}a respectively. The OCTOPUS aims at calculating $\beta=\bm{\gamma\cdot\alpha}$. The input signal has $N$ paths, and the amplitude at the $i$th path is set to be $\alpha_i$, so it can simply be denoted with $\bm{v_\alpha}=\bm{\alpha}$.

At part I, $\bm{\gamma}$ is encoded with a set of attenuators and phase shifters at each path. At the $i$th path, the attenuator controls the magnitude of $\gamma_i$, while the phase shifter determines the sign of $\gamma_i$ ( when $\gamma_i$ is negative, the phase shifter performs a $\pi$ shift on the input signal).
The total transformation can be represented by the diagonal matrix $\Gamma={\rm diag}(\gamma_1,\gamma_2,\cdots,\gamma_N)$.
So the output of part I can be represented by  
\begin{equation}
\bm{v}_1=\Gamma\bm{v_\alpha}= [ \gamma_1\alpha_1, \gamma_2\alpha_2, \cdots, \gamma_N\alpha_N]^T. \label{eq:ov}
\end{equation}

At Part II, all elements in Eq.~\eqref{eq:ov} are ``summed up" with an $n$-layer interferometer tree. The $j$th layer of the tree contains $2^{n-j}$ interferometers, each of which performs the Hadamard transformation on two nearest neighbour paths. If we denote the input of an interferometer as $[v^{\rm in}_1,v^{\rm in}_2]^T$, and the output as $[v^{\rm out}_-,v^{\rm out}_+]^T$, each interferometer performs the Hadamard transformation~\cite{Clements.16} at two nearest neighbour paths
\begin{equation}
 \begin{bmatrix}  v^{\rm out}_-\\ v^{\rm out}_+
\end{bmatrix} =H\begin{bmatrix}  v^{\rm in}_1\\ v^{\rm in}_2
\end{bmatrix} = \frac{1}{\sqrt{2}}\begin{bmatrix}  1&-1\\ 
1&1
\end{bmatrix} \begin{bmatrix}  v^{\rm in}_1\\ v^{\rm in}_2
\end{bmatrix}.\label{eq:mz}
\end{equation}
We denote the input of the $j$th layer of the tree as $\bm{v_j}$, and separate the output paths at this layer into two groups. The ``$+$" group contains all output paths corresponding to $v^+_{\rm out}$, while the ``$-$" group contains those corresponding to $v^-_{\rm out}$. They can be represented by $T_{n-j+1}^{+}\bm{v_{j}}$ and $T_{n-j+1}^{-}\bm{v_{j}}$ respectively, where $T_i^{+}$ and $T_i^{-}$ are the following $2^{i}\times2^{i-1}$ matrices:
\begin{subequations}

\begin{align}
T_{i}^{+}=\frac{1}{\sqrt{2}} \begin{bmatrix} 1&1 &&&&&&\\ 
&&1&1&&&&\\ &&&&\ddots&&&\\ &&&&&&1&1  \end{bmatrix}.\label{eq:Tj}
\end{align} 
\begin{align}
T_{i}^{-}=\frac{1}{\sqrt{2}} \begin{bmatrix} 1&-1 &&&&&&\\ 
&&1&-1&&&&\\ &&&&\ddots&&&\\ &&&&&&1&-1  \end{bmatrix}.\label{eq:Tj-}
\end{align} 
\end{subequations}

To obtain the summation, we discard the ``$-$" group, and take ``$+$" group as the input of the next [$(j+1)$th] layer 
\begin{equation}
\bm{v_{j+1}}= T_{n-j+1}^{+}\bm{v_{j}}.
\end{equation}
 So the total transformation at part II can be represented by: 
\begin{align}
\bm{v_{\rm out}}=T_1T_2\cdots T_n\bm{v_{1}}
=\frac{1}{\sqrt{N}}\sum_{j=1}^{N}\gamma_j \alpha_j=\frac{1}{\sqrt{N}}\beta,
\end{align}  
which is just the dot-product result $\beta$ multiplying a constant $\frac{1}{\sqrt{N}}$.

\textbf{Amplitude measurement.} To obtain the optical computation results, one should extract both the magnitude and the sign of the output. Here, we provide a measurement scheme in Fig.~\ref{fig:orc}b. Suppose the amplitude of the signal to be measured is $\beta$, firstly, the signal is splitted into two paths with amplitude $\beta_{\rm m}$ for main path and $\beta_{\rm anc}$ for ancillary path, where $|\beta_{\rm anc}|\ll |\beta_{\rm m}|$ is required. By measuring the intensity at the main path, one obtains the magnitude of $\beta$. Then, we introduce a reference path with signal amplitude $\beta_{\rm ref}$, and interfere it with the ancillary path. We require that $\beta_{\rm ref}\gtrsim \beta_{\rm anc}$. The ancillary path and the reference path are interfered with a $50/50$ beam splitter. Obviously, if the intensity is enhanced after the interference, the sign of $\beta$ should be ``$+$", otherwise the sign should be ``$-$".

\textbf{TLD-ONN structure.}
As can be seen in Fig.~\ref{fig:net},
 if there are $N_i$ neurons and $N_{i+1}$ neurons at the $i$th and $(i+1)$th layer, there are totally $N_i\times N_{i+1}$ output paths from the $i$th layer. Actually, the signal amplitudes of output paths coming from the same neuron (totally $N_{i+1}$ paths) are identical. So we represent the output signals from the $i$th layer with an $N_i$ dimensional vector $\bm{h^{(i)}}=\left[h^{(i)}_1,h^{(i)}_2,\cdots,h^{(i)}_{N_i}\right]^T$, where $h^{(i)}_j$ represents the amplitude of all paths output from the $j$th neuron of the $i$th layer. Note that the input of the network is just $\bm{h^{(0)}}$. 

To introduce the bias at each layer, the input of the $(i+1)$th layer is the output of the $i$th layer adding an ancillary path (at the top of each layer in Fig.~\ref{fig:net}a) with constant amplitude. Without loss of generality, we assume this constant to be 1. Therefore, there are $(N_i+1)$ input paths for each neuron at the $(i+1)$th layer, which we represent with $\bm{h'^{(i)}}=\left[h^{(i)}_1,h^{(i)}_2,\cdots,h^{(i)}_{N_i},1\right]^T$.
 
 Each neuron contains an OCTOPUS encoding a set of parameters at its part I. We denote all parameters corresponding to the $j$th neuron of the $(i+1)$th layer as 
 \begin{equation}
 W^{(i+1)}_j=\left[W^{(i+1)}_{j,1},W^{(i+1)}_{j,2},\cdots,W^{(i+1)}_{j,N_{i}},W^{(i+1)}_{j,N_{i}+1}\right]. \label{eq:wij}
 \end{equation}
 Here, $W^{(i+1)}_{j,k}$ denotes the parameter encoded at the $(i+1)$th layer, the $k$th path of the $j$th neuron, and $W^{(i+1)}_{j,N_{i}+1}$ is the bias of the $(i+1)$th layer. 

As illustrated in Fig.\ref{fig:net}b, c for TLD-ONN, at each neuron of the $(i+1)$th layer, the signal is first pass through OCTOPUS, and then a nonlinear activation $f(x)$ (see Supplemental Materials and Fig.~S1), and is finally distributed uniformly to $N_{i+2}$ paths. We define the linear transformation matrix from the $i$th layer to the $(i+1)$th layer as  
 \begin{equation}
 W^{(i)}=\left[W^{(i)}_1,W^{(i)}_2,\cdots,W^{(i)}_{N_{i+1}}\right]^T. \label{eq:Wi}
 \end{equation}
 With a little thought, one can find that  the relation between $\bm{h^{(i+1)}}$ and $\bm{h^{(i)}}$ are given by
\begin{equation}
\bm{h^{(i+1)}}= F \left(W^{(i)}\bm{h'^{(i)}}  \right), \label{eq:layer}
\end{equation}
 where $F(x)=\frac{1}{\sqrt{N_{i+2}}}f\left(\frac{1}{\sqrt{N_{i}}}x\right)$ is the rescaled nonlinear activation function. 
 As can be seen, the transformation at TLD-ONN is equivalent to the standard feed-forward neural network. The training process is discussed in the Supplemental Material~\cite{sm}. 

\textbf{RLD-ONN structure.}
The general structure of RLD-ONN is similar to LD-ONN (Fig.~\ref{fig:net}a). The difference lies in the transformation performed by each neuron. As shown in Fig.~\ref{fig:net}b,~c, each neuron contains three parts: encoding, recursively connecting, and nonlinear activation. The encoding part is identical to part I of OCTOPUS, and the nonlinear activation is discussed in Supplemental Material~\cite{sm}. So we focus on the recursively connecting part in the middle. Generally, the goal of this part is to make all input connect to all output ports while maintaining the logarithmic circuit depth, and does not discard any paths. It turns out that this can be realized with a set of revised interferometer tree.

 We recall from the Part II of OCTOPUS that at the $j$th layer of the interferometer tree, its input paths are represented with $\bm{v_{j}}$, and the output  paths are separated into ``$+$" group and ``$-$" group. As shown in Fig.~\ref{fig:net}b, c, the ``$+$" group, as usual, serves as the input of the next layer. But instead of discarding the ``$-$" groups, they are also traced. For an $n$-layer interferometer tree with input signal $\bm{v_1}$, we denote its output at the $j$th layer with $S^n_j(\bm{v_1})$, which is given by 
\begin{equation}
S_j^n(\bm{v_1})=
\left\{\begin{aligned}
&T_{n-j+1}^-\bm{v_{j}},&&j=1,\cdots,n-1\\
& H\bm{v_{n}},&&j=n.
\end{aligned} \right. \label{eq:tree}
\end{equation}
For $j<n$, the output is just the ``$-$" group at the corresponding layer; and for $j=n$, the output consists both ``$+$" and ``$-$" groups.

Obviously, only the output ports at the $n$ layer $S^n_n(\bm{v_1})$ are connected to all input ports. Signals from the remaining output ports only contain local information of the input they are connected to, so the network is not expected to work well if one use them directly. The key idea of the ``recursive" is that the interferometer trees described by Eq.~\eqref{eq:tree} are recursively appended until all input and output ports are fully connected. Take Fig.~\ref{fig:net}c as an example. We first apply a $3$-layer interferometer tree, after which the first and second paths are connected to all input ports. We then apply a $1$-layer tree to the 3rd and 4th paths, and $2$-layer tree to 5th-8th paths, after which only the A and B paths do not connect to all input ports. So we finally apply $1$-layer tree to A and B paths. Then, all output and input ports are fully connected. 
Formally, for input vector $\bm{x}$ of dimension $N=2^n$, the transformation performed by the recursive structure $\bm{y}=U(\bm{x})$ is described by  Alg.~\ref{alg:recurse}. Moreover, the training method for RLD-ONN is discussed in details in~\cite{sm}.

\begin{algorithm} [!htbp]\label{alg:train}
\caption{$U(\bm{x})$} 
\label{alg:recurse}
\begin{algorithmic}
\REQUIRE input vector $\bm{x}=[x_1,x_2,\cdots,x_N]^T$ satisfying $N=2^{n}$ with $n\in\mathbb{Z^+}$.

\STATE set $\bm{y}\leftarrow S^n_n (\bm{x})$

\STATE \textbf{if} $n>1$

\STATE \quad\textbf{for} {$i=1$, $n-1$} \textbf{do}
\STATE \quad\quad $\bm{x'}=S^n_{n-i}\bm{x}$

\STATE \quad\quad $\bm{y'}=U(\bm{x'})$

\STATE \quad\quad $\bm{y}\leftarrow [\bm{y}^T,\bm{y'}^T]^T$

\STATE \quad \textbf{end for}

\STATE \textbf{end if}

\STATE output $\bm{y}$

\end{algorithmic} 
\end{algorithm}

\textbf{Technical details for simulations.}
    In the simulation, the errors are introduced to attenuators and interferometers unless specified. For attenuator encoding the value $w$, the error is introduced by making the replacement $w\rightarrow w+\delta w$; for each interferometer, the replacement is 
\begin{equation}
H\rightarrow\widetilde{H}=\frac{1}{\sqrt{2}}\begin{bmatrix}  1+\delta&-1+\delta\\ 
1-\delta&1+\delta
\end{bmatrix}. \label{eq:mz}
\end{equation}  
 In general, the error terms should be complex values. But in Supplemental Material~\cite{sm}, we show that when $|\delta w| \ll w$ and $|\delta| \ll 1$, the effect of the imaginary parts of $\delta w$ and $\delta$ are an order smaller than the real part. So the noise effect can be well approximated by restricting $\delta w$ and $\delta$ to be real. We further assume that $\delta$ and $\delta w$ are Gaussian noises drawn from $\mathcal{N}(0,\sigma_I^2)$ and $\mathcal{N}(0,w^2\sigma_A^2)$ respectively, where $\sigma_I$ and $\sigma_A$ are noise level for interferometers and attenuators respectively.  
 
    In the simulation of linear perceptron (Fig.~\ref{fig:com}b), the input data of ``Iris" data set are four-dimensional vectors describing different properties of a particular ``Iris" flower. We test our algorithm on two kinds of Iris, ``Setos" and ``Versicolour", and label them with ``1" and ``0" respectively. Totally 100 pairs of data and labels are used, which are separated to training set (with size 60) and testing set (with size 40). 

    In the simulation of deep ONN (Fig.~\ref{fig:net}d), the networks contain one hidden layer with 64 neurons, and we assume that the activation function is inverse square root unit (ISRU), which is realized by optical amplifier (see Supplemental Material~\cite{sm}). For a given Letter, there are totally 16 primitive numerical attributes, so the input data is $16$ dimensional. Each elements of the input data are rescaled to the interval $[0,1]$. The dimension of output vectors are four, and the desired output for each letter is $\bm{y_a}=[1,0,0,0]^T$ for ``A", $\bm{y_b}=[0,1,0,0]^T$ for ``B", $\bm{y_c}=[0,0,1,0]^T$ for ``C" and $\bm{y_d}=[0,0,0,1]^T$ for ``D" respectively. We use in total 2880 pairs of inputs and labels, which are separated to training set (with size 1920) and testing set (with size 960). 

The training process follows Algorithm.S1 and Algorithm.S2 in~\cite{sm}. For SVD approach, after obtaining the well-trained $W^{(i)}$, we append it with enough ``$0$" elements until they become square matrices, and the hidden layers, input, output vectors are appended with ``$0$" accordingly. Singular-Value decomposition is performed at the matrices $W^{(i)}$, after which the parameters for attenuators and the M-Z interferometers are be determined with the method given in Ref~\cite{Clements.16}. To make the comparison fair enough, we set the learning rate ($0.01$), mini-batch size ($32$) and initial guess of all trainable parameter (drawn from $[-0.1,0.1]$ with uniform probability) to be the same for all approaches. The results shown in Fig.~\ref{fig:net}d are averaged over 10 runs with different generated random noise.


\begin{thebibliography}{42}
\bibitem [{not()}]{note_1}%
  \BibitemOpen
  \href@noop {} {}\bibinfo {note} {Since the last four output paths are
  untraced, many parts of the circuit are redundant. We still keep them in our
  illustration, in order to provide a more clear picture of the general idea of
  the circuit.}\BibitemShut {Stop}%
\end{thebibliography}
%




\vspace{1cm}
\onecolumngrid

\begin{center}
{\bf\large Supplementary material}
\end{center}
\vspace{0.5cm}

\setcounter{secnumdepth}{3}  
\setcounter{equation}{0}
\setcounter{figure}{0}
\setcounter{table}{0}
\setcounter{algorithm}{0}
\renewcommand{\theequation}{S-\arabic{equation}}
\renewcommand{\thefigure}{S\arabic{figure}}
\renewcommand{\thetable}{S-\Roman{table}}
\renewcommand{\thealgorithm}{S\arabic{algorithm}}

\renewcommand\figurename{Supplementary Figure}
\renewcommand\tablename{Supplementary Table}

\newcommand\Scite[1]{[S\citealp{#1}]}

\newcolumntype{M}[1]{>{\centering\arraybackslash}m{#1}}
\newcolumntype{N}{@{}m{0pt}@{}}

\makeatletter \renewcommand\@biblabel[1]{[S#1]} \makeatother

\onecolumngrid

This supplemental material contains the following content. In Sec.~\ref{sec:err}, we give a theoretical estimation of the noise level for OCTOPUS based and SVD based optical linear transformation; in Sec.~\ref{sec:act}, we discuss the realization of nonlinear activation; in Sec.~\ref{sec:ld} and Sec.~\ref{sec:rld}, we discuss the training process of the TLD-ONN and RLD-ONN respectively. In Sec.~\ref{sec:rd}, we show that RLD-ONN can reduce to TLD-ONN and therefore, it is universal.

\section{Error estimation for linear transformation}\label{sec:err}
Suppose we are given an input vector $\bm{x}$ and the transformation matrix $W$, our goal is to comput $\bm{y}=W\bm{x}$ optically. We restrict that $\bm{x}\equiv[x_1,x_2,\cdots,x_N]^T$, $\bm{y}\equiv[y_1,y_2,\cdots,y_M]^T$ and the transformation matrix $W$ to be real, as it is the common scenario of machine learning applications. Ideally, the amplitude of output signal of either OCTOPUS or SVD approaches are given by $\bm{v_{\rm out}}$, which satisfies
\begin{align}
\bm{v_{\rm out}} = C\bm{y}=C W\bm{x},
\end{align}
 where $CW$ is the transformation performed by the photonic circuits. As discussed in the Method section of~\Scite{main}, for OCTOPUS approach, the constant $C=1/\sqrt{N}$; for SVD, the constant $C=1$. When the encoding error is introduced, the imperfect transformation is replaced by $\widetilde{W}$, and the above equation becomes
\begin{align}
\bm{v'_{\rm out}}=C\bm{y'}=C\widetilde{W}\bm{x}.
\end{align}
We define the error term as 
\begin{equation}
\bm{\delta y}\equiv[\delta y_1,\delta y_2,\cdots,\delta y_M]^T\equiv(\bm{y'}-\bm{y}).
\end{equation}
 It is in general a complex vector, i.e., $\delta y_j=\delta y_j^{\rm re} +i\delta y_j^{\rm im}$ with $\delta y_j^{\rm re},\delta y_j^{\rm im}$ to be real and nonzero. While the vector $\bm{v'_{\rm out}}$ is the amplitude of output signal, the computation result, $\bm{\widetilde{y}}\equiv[\widetilde{y}_1,\widetilde{y}_2,\cdots,\widetilde{y}_M]^T$, can only be estimated after the measurement (see Methods in~\Scite{main}).
If we assume the measurement process to be ideal, each element is given by 
\begin{align}
\widetilde{y}_j&=|y_j+\delta y^{\rm re}_j+i \delta y^{\rm im}_j|\notag\\
&= y_i+\delta y_i^{\rm re}+O\left(\frac{(\delta y_i^{\rm im })^2}{y_i}\right).
\end{align}
Even though the imaginary part of the error is in general comparable to the real part in real experimental implementation, when $|\delta y_i|\ll y$, the \textbf{effect of imaginary part to the final result} is negligible compared with the real part. Therefore, when estimating the noise effect, we can safely simplify the noise model as  
\begin{equation}
\widetilde{y}_j=y_j + \delta y_j, \quad \delta y_j\in \mathbb{R}.
\end{equation}
For similar reason, in the following discussion, all encoding errors are assumed to be real.

\subsection{Cosine Distance}

Without loss of generality, we can assume the error term $\bm{\delta y}$ has zero mean and variance $\sigma_y^2\equiv\text{Var}(\delta y_i)=\text{Var}(\tilde{y}_i)$. We use cosine distance between $\bm{y}$ and $\bm{\widetilde{y}}$ to quantify the effect of error. When $\bm{y}\gg \bm{\delta y}$, it can be estimated by
\begin{align}
\mathcal{D}(\bm{y},\bm{\widetilde{y}})&=1 - \frac{\bm{y}\cdot\bm{\tilde{y}}}{\|\bm{y} \|\|\bm{\tilde{y}} \|}  \notag \\
&=1 - \frac{\sum y_i\tilde{y}_i}{\sqrt{\sum_i y_i^2}\sqrt{\sum_i \tilde{y}_i^2}} \notag\\
&=1 - \left( 1-\frac{\sum y_i\delta y_i +\delta y_i^2/2)}{\|\bm{y}\|^2}    \right)\left( 1+\frac{\sum y_i \delta y_i }{\|\bm{y}\|^2}    \right) + O\left(\frac{1}{\|\bm{y}\|^{4}}\right)\notag \\
&=\frac{1}{2 \|\bm{y}\|^2} \sum \delta y_i^2 + O\left(\frac{1}{\|\bm{y}\|^{4}}\right)\notag \\
&\sim\frac{N\sigma_y^2}{2N \overline{y_i^2}} \notag \\
&\sim\frac{\sigma_y^2}{ \overline{y_i^2}}. \label{eq:cosdist}
\end{align}
In the following, we will estimate the value of $\sigma_y^2$ for OCTOPUS and SVD approaches separately.  
\subsection{Error for OCTOPUS approach}\label{sec:ipo_err}
We should first estimate the error of a single OCTOPUS. We denote the transformation performed by imperfect OCTOPUS with encoding error as $\widetilde{\beta}=\bm{\widetilde{\gamma} \cdot\alpha}$.

We introduce encoding error for attenuators at part I:
\begin{equation}
\Gamma\rightarrow \widetilde{\Gamma}={\rm diag}( \gamma_1+\delta\gamma_1, \gamma_2+\delta\gamma_2,\cdots, \gamma_N+\delta\gamma_N),
\end{equation}
and interferometers at part II:
\begin{align}
T_i\rightarrow\widetilde{T}_i=\frac{1}{\sqrt{2}} 
\begin{bmatrix}
 1+\delta^{(i)}_1&1-\delta^{(i)}_1 &&&&&&\\ 
&&1+\delta^{(i)}_2&1-\delta^{(i)}_2&&&&\\ &&&&\ddots&&&\\ &&&&&&1+\delta^{(i)}_{2^{i-1}}&1-\delta^{(i)}_{2^{i-1}}  \end{bmatrix}, 
\end{align}
 where $\delta^{(i)}_{k}$ and $\delta\gamma_j$ and are assumed to be Gaussian noise drawn from $\mathcal{N}(0, \sigma_I^2)$ and $\mathcal{N}(0,\gamma_j^2\sigma_A^2)$ independently. The total transformation at Part II is replaced by
\begin{align}
\widetilde{T}=\widetilde{T}_1\widetilde{T}_2 \cdots \widetilde{T}_{n}\label{eq:T}.
\end{align}
After some calculation, we obtain 
\begin{align} 
\widetilde{T}=\frac{1}{\sqrt{N}}\begin{bmatrix}  1+\Delta_1&1+\Delta_2, &\cdots, &1+\Delta_N  \end{bmatrix} +O(\sigma_I^2)
\end{align}
where
\begin{equation}
\Delta_j=\sum_{i=1}^{n}\delta_{j}'^{(i)},\label{eq:Delta}
\end{equation}
and
\begin{eqnarray}
\delta'^{(i)}_j= \left\{
\begin{array}{rcl}
\delta^{(i)}_{\lfloor i/2^{n-j+1}\rfloor}    &  &  \mod(i/2^{n-j},2)=0 \\[6pt]
-\delta^{(i)}_{\lfloor i/2^{n-j+1}\rfloor}    &  &  \mod(i/2^{n-j},2)=1. 
\end{array} \right. \label{eq:r}
\end{eqnarray}
So the output of part II becomes:
\begin{equation}
\bm{\widetilde{v}_n}=\widetilde{T}\widetilde{\Gamma}\bm{v_\alpha} \approx\frac{1}{\sqrt{N}} \left[\beta + \sum_{j=1}^{N}\alpha_j\delta\gamma_j+\alpha_j\gamma_j\Delta_j  \right]. \label{eq:vy}
\end{equation}
Therefore, the estimated value of $\beta$ is
\begin{equation}
\widetilde{\beta}\approx\beta + \sum_{j=1}^{N}\alpha_j\delta\gamma_j+\alpha_j\gamma_j\Delta_j. \label{eq:vbeta}
\end{equation}
Since $\delta\gamma_j$ and $\delta'^{(i)}_j$ are independent of each other, for large enough $N$ we have 
\begin{equation}
\text{Var}\left(\sum_{j=1}^{N}\alpha_j\delta\gamma_j\right)\sim \|\bm{\alpha\cdot\gamma}\|^2\sigma_A^2\sim\beta^2\sigma_A^2,
\end{equation}
and
\begin{equation}
\text{Var}\left(\sum_{j=1}^{N}\alpha_j\gamma_j\Delta_j\right) \sim  \sum_{j=1}^{N}(\alpha_j\gamma_j)^2 n\sigma_I^2  \sim n\beta^2 \sigma_I^2.\label{eq:Delta2}
\end{equation}
 Therefore, the variant of $\widetilde{\beta}$ is estimated as
\begin{equation}
\sigma_\beta^2\equiv\text{Var}\left(\widetilde{\beta}\right)\sim  \left(n\sigma_I^2+ \sigma_A^2 \right)\beta^2.\label{eq:beta_var}
\end{equation}
Since each element $y_i$ are calculated with its corresponding OCTOPUS, according to Eq.~\eqref{eq:beta_var}, the variant of $y_i$ satisfies
\begin{equation}
\sigma_y^2\sim  \left(n\sigma_I^2+ \sigma_A^2 \right)\overline{y_i^2}.
\end{equation}
So  $\mathcal{D}(\bm{y},\bm{\widetilde{y}})$ can be estimated as
\begin{align}
\mathcal{D}(\bm{y},\bm{\widetilde{y}}) \sim\frac{\sigma_y^2}{\overline{y_i^2}}= n\sigma_I^2+\sigma_A^2.
\end{align}
The error contributed from Part I does not increase as $N$ increase, while the error contributed from part II increase as $O(n)=O(\log N)$. These results are consistent with the circuit depth of both part I and part II, as well as the numerical results in Fig.3a of~\Scite{main}. 

\subsection{Error for SVD approach}\label{sec:svd_err}

To begin with, we first review how SVD approach realize arbitrary real value linear transformation with photonic  circuit of totally $2N+1$ layers. We can denote the amplitudes at the $j$th layer and $i$th path as $x^{(j)}_i$, and use $\bm{x^{(j)}}=[x^{(j)}_1,x^{(j)} _2, \cdots, x^{(j)}_N]^T$ to represent the signal at the $j$th layer. The input ports and output ports of the ONN correspond to $\bm{x^{(0)}}=\bm{x}\notag$ and $\bm{x^{(2N+1)}}=\bm{y}$ respectively.

We use $M^{(j)}$ to represent the transformation at the $j$th layer, so the transformation from $\bm{x^{(j)}}$ to $\bm{x^{(j+1)}}$ can be denoted as 
\begin{equation}
\bm{x^{(j)}}=M^{(j)} \bm{x^{(j-1)}}.
\end{equation}

While $M^{(N+1)}$ is a diagonal matrix, $M^{(j\neq N+1)}$ are unitary matrices containing only the interaction between nearest neighbour paths. The interaction between nearest neighbor paths are introduced with M-Z interferometers. Since we restrict all transformation to be real, they can be represented as 
\begin{equation}
R(\theta)=\begin{bmatrix} \cos\theta& -\sin\theta\\
\sin\theta & \cos\theta
\end{bmatrix}.
\end{equation}
 We denotes $R_i(\theta)$ as the transformation performed at the $i$th and $(i+1)$th paths while all other paths remain unchanged. If we assume $N$ is even, the transformation at $j$th layer can be represented by  
\begin{eqnarray}
M^{(j)}=\left\{\begin{aligned}
\;\; \prod_{k=1}^{N/2}R_k(\theta_{2k-1,j})&&  \quad \;\; j\in\{1,3,\cdots,N-1\}\cup \{N+2,N+4, \cdots,2N\}\\
\prod_{k=1}^{N/2-1}R_k(\theta_{2k,j})&& \quad j\in\{2,4,\cdots,N\}\cup \{N+3,N+5, \cdots,2N+1\}\\
\text{Diag} \left(s_1,s_2,\cdots,s_N\right)&&   \quad\quad j=N+1\\
\end{aligned} \right. \label{eq:svd}
\end{eqnarray} 
Here, $R_k(\theta_{2k-1,j})$ and $R_k(\theta_{2k,j})$ corresponds to the M-Z interferometer at the $j$th layer and connects the $k$th and $(k+1)$th paths; $s_j\in[-1,1]$ corresponds to the attenuator at the $(N+1)$th layer and the $j$th paths.

The total transformation is given by  
\begin{equation}
W=\prod_{j=0}^{2N}M^{(2N+1-j)}.
\end{equation}
It has been shown that arbitrary real $W$ can be realized by choosing $\theta_{i,j}$ and $s_{i}$ appropriately~\Scite{sClements.16}. 

 To study the effect of encoding error, we can also do the following replacements 
 \begin{subequations}
 \begin{align}
 \theta_{i,j}&\leftarrow \theta_{i,j}+\delta\theta_{i,j},\\
 s_i&\leftarrow s_i+\delta s_i.
 \end{align}
 \end{subequations}
Similarly, we assume that $\delta\theta_{i,j}$ and $\delta s_i$ are real, and are drawn from normal distribution $\mathcal{N}(0,\sigma_I^2)$ and $\mathcal{N}(0,s_i^2\sigma_A^2)$ respectively. The transformation at each layer then becomes:
\begin{equation}
\widetilde{M}^{(j)}=M^{(j)} +\delta M^{(j)}
\end{equation}
where
\begin{eqnarray}
\delta M^{(j)}=
\left\{\begin{aligned}
&\sum _{k=1}^{N/2}\delta\theta_{2k-1,j}\frac{\partial M^{(j)}}{\partial \theta_{2k-1,j}} +O(\delta\theta_{2k-1,j}^2)  && j\in\{1,3,\cdots,N-1\}\cup \{N+2,N+4, \cdots,2N\}\\
&\sum _{k=1}^{N/2}\delta\theta_{2k,j}\frac{\partial M^{(j)}}{\partial \theta_{2k,j}}+O(\delta\theta_{2k,j}^2)&&j\in\{2,4,\cdots,N\}\cup \{N+3,N+5, \cdots,2N+1\}\\
&\text{Diag}(\delta s_1,\delta s_2,\cdots,\delta s_N) &&j= N+1.
\end{aligned} \right.\label{eq:t1}
\end{eqnarray}

The final output then becomes

\begin{align}
\bm{\widetilde{y}} &= \prod_{j=0}^{2N} \widetilde{M}^{(2N+1-j)} \bm{x}\notag\\
&= \bm{y}+\sum_{j=1}^{2N+1} \bm{\delta x^{(j)}}\notag\\
&=\bm{y}+ \bm{\delta y},
\end{align}
where
\begin{align}
\bm{\delta x^{(j)}} =  M^{(2N+1)}M^{(2N)} \cdots \delta M^{(j)} \cdots M^{(2)}M^{(1)} \bm{x}.\label{eq:t2}
\end{align}

Since $M^{(i)}$ are either unitary matrix or diagonal matrix with the values restricted to $[-1,1]$, the order of $\bm{\delta x^{(j)}}$ is determined by $\delta M^{(j)}$. We recall that the variance of $\delta\theta_{i,j}$ and $\delta s_i$ are $\sigma_I^2$ and $s_i^2\sigma_A^2$ respectively. Let $\delta x^{(j)}_i$ to be the $i$th element of $\bm{\delta x^{(j)}}$, we have 
\begin{subequations}
\begin{align}
\text{Var} \left(\delta x_{i}^{(j\neq N+1)}\right)&\sim\sigma_I^2 ,\\
\text{Var} \left(\delta x_{i}^{(j=N+1)}\right)&\sim s_i^2\sigma_A^2\sim\sigma_A^2.
\end{align}
\end{subequations}
Since $\delta y_{i}=\sum_{j=1}^{2N+1}\delta x^{(j)}_i$ and $\delta x^{(j)}_i$ are independent of each other, for large $N$ we have 
\begin{equation}
\sigma_y^2\equiv\text{Var}(\delta y_{i})\sim (2N\sigma_I^2+ \sigma_A^2).
\end{equation}
Combining with Eq.~\eqref{eq:cosdist}, the cosine distance can be estimated as  
\begin{equation}
\mathcal{D}(\bm{y},\bm{\widetilde{y}})=\frac{\sigma_y^2}{\overline{y_i^2}}\sim (N\sigma_I^2+ \sigma_A^2).
\end{equation}
So $\mathcal{D}(\bm{y},\bm{\widetilde{y}})$ increases as $O(N)$, which agrees well to the linear depth of SVD circuit and the numerical result in Fig.3a of~\Scite{main}. 

\section{ Amplifier as activation function}\label{sec:act}

\begin{figure} [!htb]
\includegraphics[width=0.55\columnwidth]{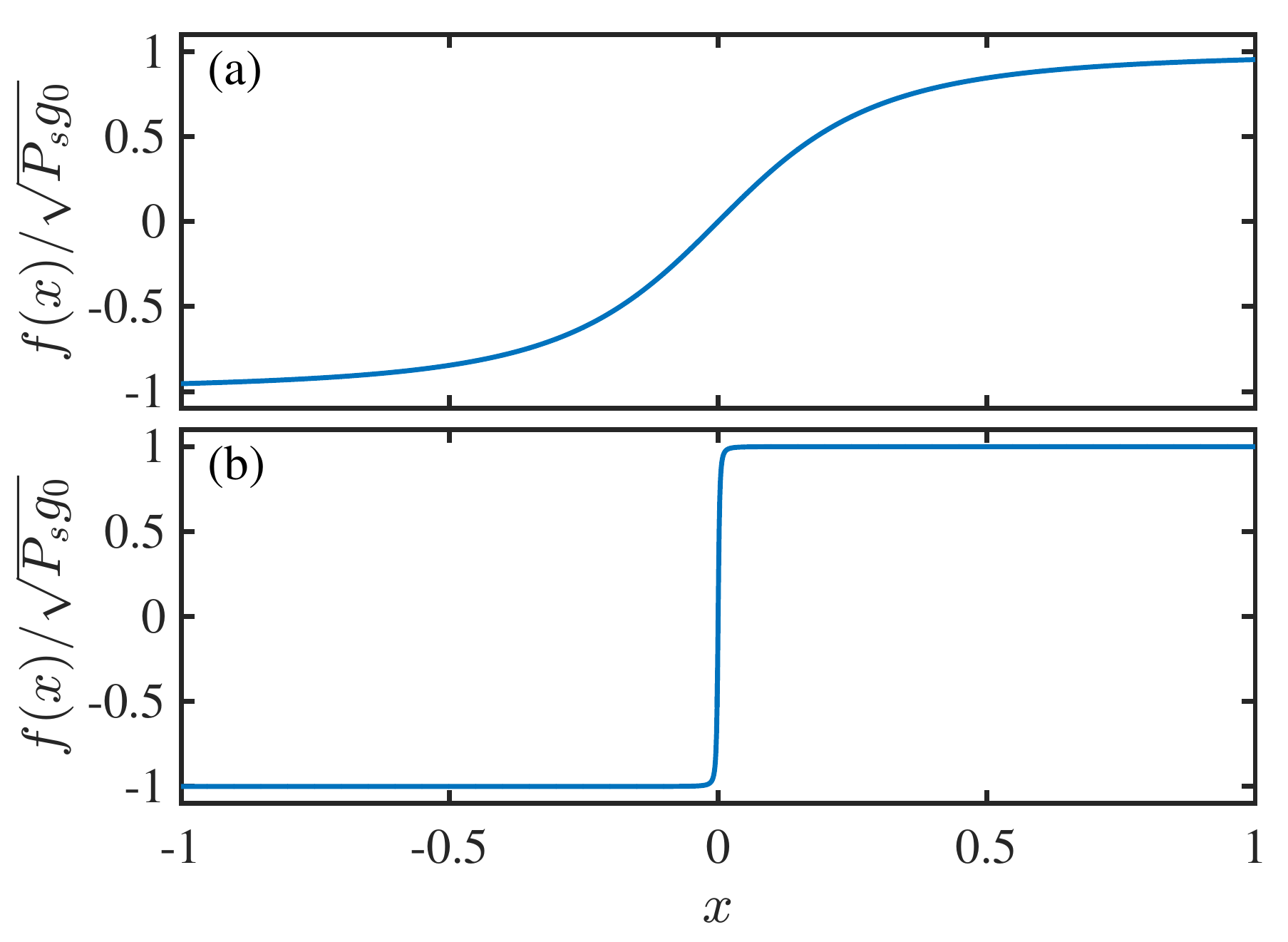}
\caption{ Optical amplifier as activation function. x-axis (y-axis) is the input (output) signal. Saturation power is set to be $P_{s}=10^{-1}$ for (a) and $P_{s}=10^{-5}$ for (b), corresponding to ISRU and binarized function respectively.
}
\label{fig:act}
\end{figure}

There are many ways to generate different types of nonlinear activation, such as with the saturable absorbers~\Scite{sShen.17} or via measurement~\Scite{sHughes.18}. With the former one, the transformation is realized with the speed of light, but the signal will attenuate when passing through each layer. With the measurement approach, the intensity will not decrease as the network size increase, but the computation speed would be reduced by the measurement process. Here, we introduce an alternative approach based on optical amplifier, which does not suffer from signal strength reduction, and maintain the high computation speed. But the most appropriate solution to the activation function depends on the practical scenarios, and require further studies.

For an optical amplifier, the power of input and output signals are given by $P_{\rm in}=|x|^2$ and $P_{\rm out}=|f(x)|^2$ respectively, where $x$ and $f(x)$ are the amplitude of input and output signal. The gain of an optical amplifier, $g=P_{\rm out}/P_{\rm in}$, generally satisfies~\Scite{sConnelly.07}:
\begin{equation}
g=\frac{g_0}{1+P_{\rm in}/P_{\rm s}}.
\end{equation}
where $P_s$ is the saturation power. So for an amplifier that maintains the signal phase, the nonlinear relation between input and output is 
\begin{align}
f(x)=\frac{\sqrt{g}_0 x}{\sqrt{1+x^2/P_{s}}},\label{eq:isru}
\end{align}
which is exactly the inverse square root unit (ISRU)~\Scite{Carlile.17}. Through out our simulation, Eq.~\eqref{eq:isru} is used as the activation function. On the other hand, if the amplifier works in the saturation regime, i.e. $x\ll \sqrt{P_s}$, we have the following binarized activation function
\begin{align}
f(x)=  (g_0P_s)^{1/2} \text{sgn}(x),\label{eq:bi}
\end{align}
with which one can construct a binarized neural network~\Scite{Courbariaux.16,Hubara.16}. This type of neural networks only need the amplitudes $\bm{h^{(i)}}$ and the elements of matrices $W^{(i)}$ to be either $+1$ or $-1$, but still have comparable performance to traditional networks. So one may encode the parameters only without tunnable attenuator, and dramatically simplify the structures. 

In Fig.~\ref{fig:act}, we show the input-output relations for different values of $P_s$, corresponding to Eq.~\eqref{eq:isru} and Eq.~\eqref{eq:bi} respectively.

\begin{algorithm} [!htbp]
\caption{Training for TLD-ONN} 
\label{alg2} 
\begin{algorithmic}

\STATE Initialize $W^{(i)}$ for $i=1,2,\cdots L$, and set learning rate $\alpha$

\FOR{iteration $=1$, $N_{\rm iter}$} 
\STATE \quad Sample a minibach of input data and labels $\mathcal{B}$ with size $N_b$

\quad Set $\bm{\delta^{(L)}}=0$

\quad\textbf{for} $i_{\rm b}=1$, $N_{\rm b}$ \textbf{do}    

\quad \quad Set $\bm{h^{(0)}}=\bm{x_{i_b}}$

\quad \quad\textbf{for} $i=1$, $L-1$, \textbf{do}   

\quad  \quad \quad Calculate $\bm{z^{(i)} }=W^{(i)}\bm{h'^{(i)}}$ and store $\bm{z^{(i)}}$ $\quad\#$ can be realized with OCTOPUS

\quad  \quad \quad Calculate $\bm{h^{(i+1)}} = F (\bm{z^{(i)}})$ and store $\bm{h^{(i+1)}}$ 

\quad \quad \textbf{end for}

 
  \quad \quad $\bm{\delta^{(L)}} = \bm{\delta^{(L)}}+ \frac{1}{N_b}\frac{\partial\mathcal{L}}{\partial{\bm{h^{(L)}}}}  \odot F'\left(\bm{z}^{(i)}\right)$ \Scite{pd}

 \quad\textbf{end for}
 

\quad\textbf{for} $i=1$, $L-2$, \textbf{do}   

\quad \quad Calculate $\bm{\eta^{(L-i+1)}}=\left(W^{(L-i+1)} \right)^T \bm{\delta^{(L-i+1)}}$ $\quad\#$ can be realized with OCTOPUS

\quad \quad Calculate $\bm{\delta^{(L-i)}}=\bm{\eta^{(L-i+1)}}\odot F'\left(\bm{z^{(L-i+1)} }\right) $, and store $\bm{\delta^{(L-i)}}$

\quad\textbf{end for}

\STATE \quad Update $W^{(i)}_{j,k} = W^{(i)}_{j,k} - \alpha \delta^{(i)}_{j} h'^{(i-1)}_{k}$

\ENDFOR 

\end{algorithmic} \label{alg:ld}
\end{algorithm}

\begin{algorithm} [!htbp]
\caption{Training for RLD-ONN} 
\begin{algorithmic}

\STATE Initialize $\Theta$, and set learning rate $\alpha$

\FOR{iteration $=1$, $T$} 
\STATE \quad Sample a minibach $\mathcal{B}$ with size $N_b$

\quad Set $\delta w^{(i)}_{j,k}=0$ for all $i,j,k$

\quad\textbf{for} $i_b=1$, $N_b$ \textbf{do}    

\quad\quad\textbf{for} all possible values of  $\{i,j,k\}$ \textbf{do}

\quad\quad\quad Estimate $\frac{\partial \mathcal{L}}{\partial w^{(i)}_{j,k}}$ with Eq.~\eqref{eq:grad}

\quad\quad\quad Set $\delta w^{(i)}_{j,k} \leftarrow \delta w^{(i)}_{j,k} + \frac{1}{N_b}\frac{\partial \mathcal{L}}{\partial w^{(i)}_{j,k}}\cdot \alpha$

 \quad\quad\textbf{end for}

 \quad\textbf{end for}
 
 \quad\textbf{for} all possible values of  $\{i,j,k\}$ \textbf{do}    
 
\STATE \quad\quad Update $w^{(i)}_{j,k} \leftarrow w^{(i)}_{j,k} - \delta w^{(i)}_{j,k}$

 \quad\textbf{end for}

\ENDFOR 

\end{algorithmic} \label{alg:rld}
\end{algorithm}

\section{Training for TLD-ONN}\label{sec:ld}
The discrepancy between desired output (or label), $\bm{y}$ and the output of neural network $\bm{h^{(L)}}$ is quantified by the loss function $\mathcal{L}$, which is taken to be the mean square error in this work
\begin{equation}
\mathcal{L}=\frac{1}{2} \left \|\bm{\bm{y}}-\bm{h^{(L)}} \right\|^2. \label{eq:loss}
\end{equation}

During training, our goal is to minimize the loss function for the given training set by tuning $W^{(i)}$ containing the weights and biases of the $i$th layer. The process follows the standard back-propagation method as shown in Algorithm.~\ref{alg:ld}, where we randomly generate a minibatch containing $N_b$ pairs of input data and labels $(\bm{x_{i_b}},y_{i_b})\in\mathcal{B}$ at each iteration. The training can be realized solely in the electronic devices, after which the well-trained parameters are encoded to the ONN setup. Alternatively, one may train the network with the assistant of OCTOPUS: by executing all linear transformation steps (commented steps) with OCTOPUS, the training can be accelerated dramatically. 

We also note that it is possible to implement the ``forward propagation" approach, which train the ONN directly by tuning the attenuators and phase shifters, and obtain the gradient by directly measure the output of the ONN~\Scite{sShen.17}. 

\section{Training for RLD-ONN}\label{sec:rld}

Since RLD-ONN has a special structure, standard back-propagation is no-longer available. Instead, one can train the RLD-ONN with ``forward-propagation" method, i.e., perturb each parameter directly, and update the network according to the gradient of the cost function with respected to all trainable parameters~\Scite{sShen.17}.

Similar to TLD-ONN, we denote $w^{(i)}_{j,k}$ as the trainable parameter (encoded at the attenuators and phase shifters) at the $i$th layer, the $j$th neuron, and the $k$th path. We simply use $\Theta$ to represent all parameters of the neural network, and use $G(\bm{x},\Theta)$ to represent the output of the neural network with input $\bm{x}$ and parameter $\Theta$. $G(\bm{x},\Theta)$ can be calculated in an electronic device, or it can be estimated directly at RLD-ONN. The loss function [Eq.~\eqref{eq:loss}] then becomes
\begin{equation}
\mathcal{L}(\bm{x},\Theta)=\frac{1}{2} \left \|\bm{y}- G\left(\bm{x}, \Theta \right) \right\|^2. 
\end{equation}
We further define the perturbed output  the neural network $G\left(\bm{x}, \Theta,w^{(i)}_{j,k},\delta\right)$, which is the neural network output with parameter $\Theta$, except for the element $w^{(i)}_{j,k}$ changed as  $w^{(i)}_{j,k} \leftarrow w^{(i)}_{j,k}+\delta$. The derivative of the loss function with respected to $w^{(i)}_{j,k}$ can be estimated with
\begin{equation}
\frac{\partial \mathcal{L}}{\partial w^{(i)}_{j,k}} \simeq \left[ G\left(\bm{x}, \Theta \right) - \bm{y}\right] \frac{G\left(\bm{x}, \Theta,w^{(i)}_{j,k},\delta\right)-G\left(\bm{x}, \Theta\right)}{\delta}, \label{eq:grad}
\end{equation}
 where we assume $\delta\ll w^{(i)}_{j,k}$. $\Theta$ is updated according to the gradient of the loss function. At each iteration $i_b$, we calculate the gradient with respected to a minibatch containing $N_b$ pairs of input data and label $(\bm{x_{i_b}},y_{i_b})\in\mathcal{B}$, and perform the gradient descendent base on it. The full training algorithm is given in Algorithm.~\ref{alg:rld}.

\section{Universality of RLD-ONN}\label{sec:rd}
\begin{figure} [!htb]
\includegraphics[width=0.6\columnwidth]{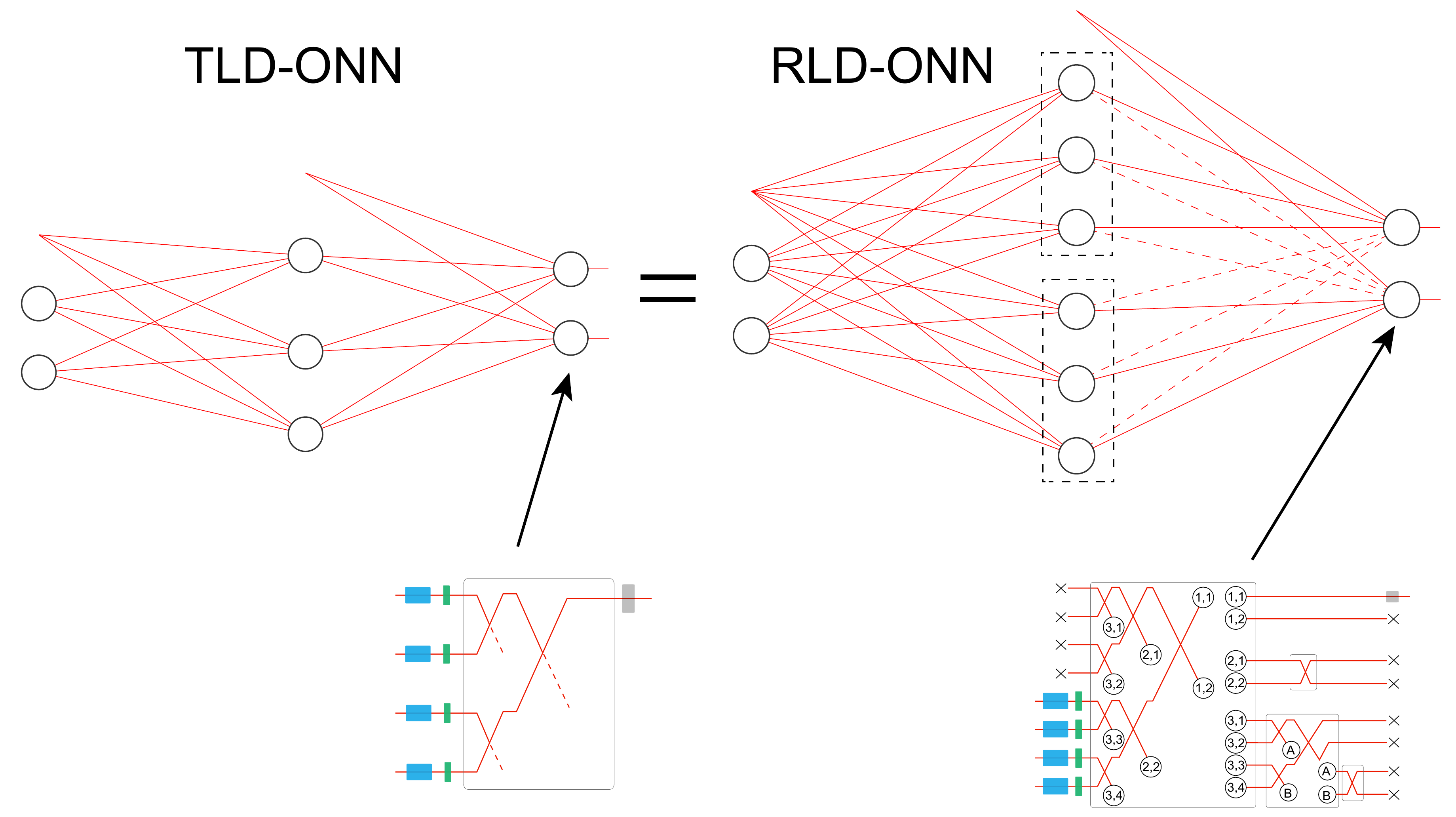}
\caption{  Equivalence of RLD-ONN and TLD-ONN.
}
\label{fig:eq}
\end{figure}


To show the universality of RLD-ONN, one just need to ensure that for TLD-ONNs with any given parameters, there are a RLD-ONN that can reduce to it. This turns out to be true. 
We consider a one-hidden-layer TLD-ONN with $N_{\text{i}}$ input, $N_{\text{h}}$ hidden, and $N_{\text{o}}$ output neurons respectively. Suppose the parameters encoded at the $i$th neuron of the hidden layer is 
\begin{equation}
\bm{W^{(h)}_i}=\left[W^{(h)}_{i,1},W^{(h)}_{i,2},\cdots,W^{(h)}_{i,N_{\text{i}}+1}\right]^T,
\end{equation}
 and the parameters encoded at the $i$th neuron of the output layer is	
 \begin{equation}
\bm{W^{(o)}_i}=\left[W^{(o)}_{i,1},W^{(o)}_{i,2},\cdots,W^{(o)}_{i,N_{\text{h}}+1}\right]^T.
\end{equation}

 In the following, we show that this TLD-ONN is identical to an RLD-ONN with $N_{\text{i}}$ input, $N_{\text{h}}\times N_{\text{o}}$ hidden, and $N_{\text{o}}$ output neurons. We denote the parameters encoded at the $i$th neuron at the hidden layer and output layer as 
\begin{equation}
\bm{V^{(h)}_i}=\left[V^{(h)}_{i,1},V^{(h)}_{i,2},\cdots,V^{(h)}_{i,N_{\text{i}}+1}\right]^T,
\end{equation}
and
\begin{equation}
\bm{V^{(o)}_i}=\left[V^{(o)}_{i,1},V^{(o)}_{i,2},\cdots,V^{(o)}_{i,(N_{\text{h}}+1)\times N_{\text{o}}}\right]^T.
\end{equation}
Firstly, we let the first $N_{\text{h}}$ hidden neurons just connect to the first output layer, the $(N_{\text{h}}+1)$ to $2N_{\text{h}}$ hidden neurons just connect tot he second output and so on. In other words, at the output layer, we set
\begin{eqnarray}
V^{(o)}_{i,j}=\left\{\begin{aligned}
\;\;0&&  \quad \;\; \quad j\leqslant(i-1)N_h \;\text{or}\; j>iN_h\\
W^{(o)}_{i,\mod(j,N_h)}&& (i-1)N_h <j\leqslant iN_h.\\
\end{aligned} \right. \label{eq:VW}
\end{eqnarray} 
As can be seen, there are totally $N$ nonzeros input at the output layers, which is the same as its corresponding TLD-ONN. Then, one just need to ensure that the input signals of the output layer are identical to those in TLD-ONN, by setting $\bm{V^{(h)}_i}$ appropriately.
For the $i$th neuron at the hidden layer, the signal before entering the nonlinear activation can be represented by
\begin{equation}
\bm{h_i}\equiv  U\left(\bm{V^{(h)}_i}\odot \bm{x'}\right) =[h_{i,1},h_{i,2},\cdots,h_{i,N_{\text{i}}+1}]^T,
\end{equation}
with $\bm{x'}=[\bm{x}^T,1]^T$ the input of ONNs. As discussed in the main text, $U$ is realized by a set of interferometers, so it corresponds to a unitary transformation. In order words, $U(\bm{x})=U\bm{x}$ for certain unitary matrix $U$. So we have 
\begin{equation}
h_{i,j}=\bm{u_j}^T ( \bm{V^{(h)}_i}\odot \bm{x'})=\left(\bm{u_j}\odot \bm{V^{(h)}_i}\right)^T \bm{x'}
\end{equation}
 for certain $\bm{u_j}=[u_{j,1},u_{j,2},\cdots,u_{j,N}]^T$. 
 As can be inferred from Eq.~\eqref{eq:VW}, only the path corresponds to $h_{i,\lceil i/N_{\text{o}}\rceil}$ are used in the next layer. Therefore, to mach the corresponding TLD-ONN, the only constrain is 
\begin{equation}
h_{i,\lceil i/N_{\text{o}}\rceil}=\bm{W^{(h)}_i}\cdot \bm{x'},
\end{equation}
which can be satisfied by setting 
\begin{equation}
\bm{V^{(h)}_i}=\bm{u_{\lceil i/N_{\text{o}}\rceil}^{-1}}\odot\bm{W^{(h)}_i},
\end{equation}
where we have defined $\bm{u_j^{-1}}=[u_{j,1}^{-1},u_{j,2}^{-1},\cdots,u_{j,N}^{-1}]^T$. An illustration of the case $N_{\text{i}}=2$,$N_{\text{h}}=3$,$N_{\text{i}}=2$ is shown in Fig.~\ref{fig:eq}.

Here, we have shown that RLD-ONN with more hidden neurons can reduce to TLD-ONN, so it is universal. But in practical application, the RLD-ONN may need much less hidden neurons than $N_{\text{h}}\times N_{\text{o}}$ to achieve a comparable performance with TLD-ONN.

\end{document}